\documentclass[usenatbib,onecolumn]{mn2e}
\def\gsim{\ \raise 3pt \hbox{$>$} \kern -8.5pt \raise -2pt \hbox{$\sim$}\ }
\def\lsim{\ \raise 3pt \hbox{$<$} \kern -8.5pt \raise -2pt \hbox{$\sim$}\ }

\newcommand{\pe}{_{\rm pe}}

\newcommand{\st}{_{\rm st}}

\newcommand{\br}{_{\rm break}}

\usepackage{amssymb}
\usepackage{graphicx}

\begin{document}
   \title[GRB spectral modeling]{GRB spectral parameters within the fireball model}

\author[G. D. Fleishman and F. A. Urtiev]{G. D. Fleishman$^{1,2}$ 
 and F. A. Urtiev$^{3}$
\\
$^{1}$New Jersey Institute of Technology, Newark, NJ 07102; \\
$^{2}$Ioffe Physical-Technical Institute of the Russian
Academy of Sciences, St. Petersburg 194021, Russia; \\
$^{3}$State Polytechnical University, St.Petersburg, 195251, Russia}


\date{Accepted 2010 March 19. Received 2010 March 17; in original form 2009 December 7}

\pagerange{\pageref{firstpage}--\pageref{lastpage}} \pubyear{2009}

\maketitle

\label{firstpage}

\begin{abstract}
Fireball model of the GRBs predicts generation of numerous internal
shocks, which then efficiently accelerate charged particles and
generate magnetic and electric fields. These fields are produced in
the form of relatively small-scale stochastic ensembles of waves,
thus, the accelerated particles diffuse in space due to interaction
with the random waves and so emit so called Diffusive Synchrotron
Radiation (DSR) in contrast to standard synchrotron radiation they
would produce in a large-scale regular magnetic fields. 
In this paper we present first results of comprehensive modeling of
the GRB spectral parameters within the fireball/internal shock
concept. We have found that the non-perturbative DSR emission
mechanism in a strong random magnetic field is consistent with
observed distributions of the Band parameters and also with
cross-correlations between them; this analysis allowed to restrict
GRB physical parameters from the requirement of consistency between
the model and observed distributions.
\end{abstract}


\section{Introduction}

The fireball model is currently accepted as a standard model of the
gamma-ray burst (GRB) prompt emission \citep[e.g.,][]{Mezaros1}. It
is supposed that a central engine produces a number of relativistic
internal shocks, which then interact with each other. The phenomenon
of the shock waves requires an efficient mechanism of energy
dissipation. In a collisionless case, the most efficient ways of the
energy dissipation are via generation of fluctuating electromagnetic
fields and acceleration of charged particles up to high energies.

Microscopically, this field generation can be driven by two-stream
instabilities associated with the shock propagation
\citep{Kazimura_1998, Medvedev_1999, Frederiksen_2004, Bret_2004,
Bret_2005, Nishikawa_2003, Jaroschek_2004, Hededel_PhD2005,
Hededal_Nishikawa_2005, Bret_Dieckmann_2008, Keshet_2008,
Dieckmann_Bret_2010}, while the acceleration of particles is
provided by their interaction with the shock-generated random and
regular electromagnetic fields \citep{Meszaros_2002,
Nishikawa_etal_2005, Piran_2005, Sari_2006, Silva_2006}. It is well
established by now that the magnetic and electric fields produced in
the shock interactions have often a significant random component at
various spatial scales.

The presence of the random component is critically important for
generation of nonthermal emission from corresponding objects.
Indeed, unlike regular gyration in the presence of a regular
magnetic field, the shock-accelerated charged particles moving
through a plasma with random electromagnetic fields experience
random Lorenz forces and so follow random trajectories representing
a kind of spatial diffusion. Accordingly, the particles produce a
diffusive radiation whose spectra depend on the type of the field
(magnetic or electric) and on spectral energy distribution of the
field over the spatial scales \citep{Topt_Fl_1987, Hededel_PhD2005,
Fl_2006a, Fl_Biet_2007, Fl_Topt_2007a, Sironi_Spitkovsky_2009}.
Below in this paper we rely on analytical DSR theory proposed by
\cite{Topt_Fl_1987} and then further developed by \cite{Fl_2006a,
Fl_Biet_2007}. An alternative way of calculating radiation is the
use of numerical PIC simulations \citep{Hededel_PhD2005,
Hededal_Nishikawa_2005, Nishikawa_etal_2008,
Sironi_Spitkovsky_2009}, which confirm the analytical results in the
common parameter domain. However, the case of strong random field
and strong angular scattering of the radiating electrons requiring a
large dynamic range of the involved parameters is yet beyond
available PIC capacities, which justifies the choice in favor of the
well tested analytical theory.

Individual spectra of the prompt GRB emission are typically well
fitted by a phenomenological Band function \citep{Band1993}, which
consists of low-energy (spectral index $\alpha$) and high-energy
(spectral index $\beta$) power-law regions smoothly linked at a
break energy $E_{br}$. The diffusive synchrotron radiation (DSR) was
shown \citep{Fl_2006a} to produce spectra consistent with that
observed typically  from the GRBs \citep{Band1993, Mazets_etal_2004,
Ohno__Palshin_etal_2008, Palshin_etal_2008, Granot_etal_2009}. It is
yet unclear, however, if the DSR spectra are naturally consistent
with observed distribution of the GRB spectral parameters
\citep{Precee_2000, Kaneko_etal_2006} and what ranges of physical
GRB parameters are needed to reconcile the theoretical spectra with
the observed ones. In this paper we present a model of GRB prompt
emission generation by DSR in relativistically expanding GRB jets.
The input parameters of the model are constrained by available
observations and take into account dependences between involved
parameters implied by physical laws. We vary a number of free
parameters of the model to achieve the best agreement between the
variety of the modeled and observed spectra. This analysis confirms
that the DSR model, specifically---the non-perturbative strong-field
regime, is intrinsically consistent with the observed distributions
of the GRB spectral parameters.

\section{Formulation of the Model}

To be specific we adopt the fireball model in which the GRB prompt
emission is generated in a collimated jet ejected with a
relativistically high speed $v$ from a central engine. Adopting a
general internal shocks/fireball concept we accept that a single
binary collision of relativistic internal shocks results in a single
episode of the GRB prompt emission. Microscopically, this
shock-shock interaction first produces high levels of random
magnetic and/or electric fields and accelerates the charged
particles up to large ultrarelativistic energies; and then, these
particles interact with the random fields to generate the
gamma-rays. Although there are some common general properties of all
cases of relativistic shock interactions, each shock-shock collision
is, nevertheless, unique in terms of combination of the physical
parameters involved. Accordingly, we are going to estimate and adopt
a set of standard ("mean") parameters appropriate to account for the
most global GRB properties, and then consider if a reasonable
scatter of those standard parameters is capable of reproducing more
detailed properties of the considered class of events as a
whole---the statistical distributions of the GRB spectral parameters
and cross-correlations between them. To do so, we consider a number
of different emission models including the standard synchrotron
radiation and DSR  regimes in case of either weak or strong random
magnetic field. The spectral slopes and breaks depend on both the
emission mechanism and combination of physical parameters affecting
the radiation spectra within a given mechanism. Thus, the goal of
the modeling is to establish if there exists a parametric space
making one or another theoretical model compatible with the
observational data on the GRB spectral properties.

\subsection{Basic parameters of the GRB source}
\label{S_basic_parms}

From analysis of so-called 'compactness problem', it is established
that the bulk Lorenz-factor of the expanding jet
$\Gamma=1/\sqrt{1-v^2/c^2}$, where $c$ is the speed of light, must
be much larger than unity, $\Gamma \gtrsim 100$ \citep{Rhoads1993,
Piran_2005, Sari_2006}. A maximum $\Gamma$ value is not well
constrained observationally; being conservative, we will not
consider values above 1000. The total kinetic energy of the jet is
roughly $E\sim 10^{51}$~erg \citep{Mezaros1}, which, along with
$\Gamma$ estimate and an assumption of particle composition, allows
estimating of the total number of ejected particles. We adopt that
bulk of the jet mass resides in the protons, thus, the total number
of ejected protons is estimated as
\begin{equation}
\label{N_chislo_chastits}N=E/(\Gamma m_p c^2)\sim 10^{51},
\end{equation}
where $m_p$ is the proton (rest) mass. If no $e^+e^-$ pairs are
produced, then the number of electrons is equal to the evaluated
number of protons.

The number density of the particles is
\begin{equation}
\label{conc_01}n=\frac{N}{V},
\end{equation}
where $V$ is the jet volume. The full volume occupied by the jet
material is a cone with the opening angle $\Omega_{ang}$ (for the
modeling we adopted a constant value of
$\Omega_{ang}=10^{-3}$~ster), so $V_{cone}=(4 \pi/ 3) \Omega_{\rm
ang} R^3$, while the volume of spherical layer with thickness
$\Delta$ of this cone is $V=(4 \pi/ 3) \Omega_{ang} (R^3-(R-
\Delta)^3)$. The volume is not well constrained by the observations,
although it can be estimated using observed time scale $\delta t$ of
the emission variability. Indeed, for a relativistically moving
source \cite[e.g.,][]{Piran_2005} we have $\delta t\approx \Delta/(2
c \Gamma^2)$; accordingly, for the co-moving frame we can estimate:
\begin{equation}
\label{volume_01} V'=\frac{4\pi}{3} \Omega_{ang}
\Delta'^3=\frac{4\pi}{3} \Omega_{ang} 8 c^3 \Gamma^3 \delta t^3.
\end{equation}
Then, the number density in the co-moving frame is
\begin{equation}
\label{n_01} n'=\frac{N}{V'}.
\end{equation}

The radiation spectra for each set of the source parameters are
calculated in the co-moving system and then the emission frequency
is transformed to the observer's frame taking into account the
relativistic motion of the source and cosmological expansion of the
Universe
\begin{equation}
\label{sdwig_omega} \omega'=\frac{\omega}{2\Gamma}(1+z).
\end{equation}

Let us introduce the energy contents \citep[e.g.,][]{Sari_2006} of
constituents needed to produce radiation---the magnetic field
\begin{equation}
\label{eB} \epsilon_B\equiv\frac{U_B}{w}=\frac{B^2}{8 \pi w},
\end{equation}
where $w=\Gamma n m_p c^2$ is the kinetic energy density of the
expanding shell and $U_B=\frac{B^2}{8 \pi}$ is the energy density of
the magnetic field; and accelerated electrons
\begin{equation}
\label{ee} \epsilon_e\equiv\frac{U_e}{w}=\frac{\gamma\xi_e
m_e}{\Gamma  m_p},
\end{equation}
where $U_e=n_e \gamma\xi_e m_e c^2$ is the energy density of
accelerated electrons, $\xi_e$ is a fraction of electrons being
accelerated ($\xi_e<1$ if only a fraction of all electrons are
accelerated \citep{Bykov_Meszaros_1996}, while others remain "cold",
and $\xi_e>1$ if $e^+e^-$ pairs are produced at the emission
source), $\gamma$ is a characteristic Lorenz-factor of the
accelerated electrons, and $m_e$ is the mass of electron.

\subsection{Standard synchrotron model}

\begin{figure}
 \centerline{
\includegraphics[height=3.0in]{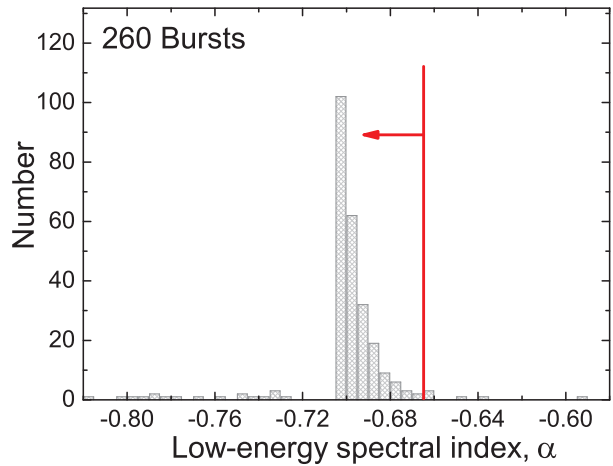}}
\caption{\small Model histogram of the spectral index distribution
assuming slow cooling synchrotron regime. Note, that most of the
indices are displaced by a small value compared with the asymptotic
value, $-2/3$ (the 'line of death'). There are a number of outliers
with a few of them apparently violating the 'line of death'. These
outliers originate from the fit errors provided that the Band
fitting function does not represent a perfect match to the
theoretical synchrotron spectrum. \label{alpha_historg}}
\end{figure}

Although synchrotron models are generally consistent with overall
GRB  energetics and light curves, they are intrinsically
incompatible with the distribution of low-energy spectral index
$\alpha$ \citep[e.g.,][]{Baring_Braby_2004}. To demonstrate this
explicitly, we show the model distribution of the low-energy indices
obtained within a synchrotron model in Figure~\ref{alpha_historg}.

To be specific in generating this histogram we assumed the slow
cooling regime and reasonable statistical distributions of the
relevant parameters (see below for greater detail), such as bulk
Lorentz factor of the relativistically expanding shell, energy
content of the accelerated particles and produced magnetic field,
number density of the particles, and their energy spectrum, as well
as correlations between these parameters consistent with
observations \citep{Mezaros1, Sari_2006}. The  histogram in
Fig.~\ref{alpha_historg}, representing an asymmetric narrow
distribution peaking around $\alpha=-0.7$, is in evident
contradiction with the observed one \citep{Precee_2000,
Baring_Braby_2004, Kaneko_etal_2006}, which is a more or less
symmetric broad distribution peaking at $\alpha \approx -1$. The
fast cooling regime results in a histogram similar to that in
Figure~\ref{alpha_historg} with the only difference that it peaks
around $\alpha\approx-1.5$. We conclude that a more sophisticated
modeling is needed to achieve reasonable agreement between the
observations and the theory of electromagnetic emission in the GRB
sources. To address this problem, Medvedev (2000) proposed that
emission of fast electrons moving in \emph{small-scale} random
magnetic field may possess the spectral properties consistent with
those observed from GRBs; the corresponding DSR process  in the GRB
context has than been studied quantitatively by \citet{Fl_2006a}
within general concept of the stochastic theory of radiation
proposed by \citet{Topt_Fl_1987}.

\subsection{Theory of DSR: main equations and parameters}

For the purpose of this more detailed modeling we note that at the
sites where the internal shocks interact in the GRB sources, charged
particles are accelerated and two-stream instabilities produce high
level of random magnetic and/or electric fields, so the diffusive
synchrotron radiation \citep{Fl_2006a} is expected to be produced
there. Let us remind basic equations describing the DSR and main
parameters determining its spectrum. We adopt that the energy of the
random magnetic field is distributed over spatial scales according a
power-law:
\begin{equation}
\label{K_alpha_beta_Furie}  K_{\alpha \beta}(q_0, \textbf{q})=
 \frac{1}{2} K(q_0,\textbf{q})\left(\delta_{\alpha\beta}-\frac{k_\alpha k_\beta}{k^2}\right)
 =\frac{1}{2} K(\textbf{q})
\delta(q_0-q_0(\textbf{q}))\left(\delta_{\alpha\beta}-\frac{k_\alpha
k_\beta}{k^2}\right),
\end{equation}
where
\begin{equation}
\label{magn_spectr} K(\textbf{q})=\frac{A_{\nu}}{q^{\nu+2}},\quad
A_{\nu}=a_{\nu}q_{\min}^{\nu-1} \langle B\st^2
\rangle=\frac{(\nu-1)q_{\min}^{\nu-1} \langle B\st^2 \rangle}{4
\pi},\quad q_{\min} < q < q_{\max},
\end{equation}
$\nu$ is the spectral index of the random field distribution, the
spectrum $K(\textbf{q})$ is normalized by $d^3q$ so that
\begin{equation}
\label{magn_spectr2}
\int_{q_{\min}}^{q_{\max}}K(\textbf{q})d^3q=\langle B\st^2 \rangle
,\quad при \quad q_{\min} \ll q_{\max}, \nu > 1,
\end{equation}
$\langle B\st^2 \rangle$ is the mean square of the random magnetic
field.

For the adopted spectrum of the random magnetic field, the
perturbative DSR spectrum  \citep{Fl_2006a} can be obtained
analytically:
\begin{equation}
\label{Intensity_q_02}
 I_{\omega}^{\bot}= \frac{8Q^2
\gamma^{2}_{*}}{3 \pi c} \cdot q(\omega),
\end{equation}
where $\gamma_{*}^{-2}=\gamma^{-2}+\omega\pe^2 / \omega^2$,
\begin{equation}
\label{Intensity_q_pereturb}
 q(\omega)= \left\{
\begin{array}{ll}
 \frac{3 \cdot 2^{\nu-1} \pi^2
 (\nu^2+3\nu+4)a_{\nu}}{\nu(\nu+2)^2(\nu+3)}
 \frac{\omega_{0}^{\nu-1} \omega\st^2 \gamma_{*}^{2\nu}}{\omega^{\nu}\gamma^2} \qquad \qquad
 \qquad \qquad\qquad\qquad \textrm{for  }
 \omega<\frac{\omega\pe^2}{2\omega_0}\ \textrm{or  }\omega
> 2\gamma^2 \omega_0 \textrm{,}\\ 
\frac{\pi^2
 a_{\nu} \omega\st^2}{2 \nu \omega_{0}  \gamma^2}
 \left[1+
 \frac{3\nu(\nu+1) \omega^2
}{4(\nu+2)^2 \gamma_{*}^{4} \omega_{0}^2}-\frac{\nu \omega^3
}{2(\nu+3)\gamma_{*}^{6}\omega_{0}^3}+\frac{3
\nu\omega^2}{4(\nu+2)\gamma_{*}^{4} \omega_{0}^2} \ln \left(
\frac{\omega}{2\gamma_{*}^{2} k_0 c}\right)\right] \ \textrm{for }
\frac{\omega\pe^2}{2\omega_0}\leq\omega\leq 2\gamma^2 \omega_0
\textrm{.}
\end{array}\right.
\end{equation}

These equations describe the DSR from particles moving along almost
rectilinear trajectories and so valid only for relatively weak
magnetic field. In the internal shock interactions, however, a
strong random magnetic field can often be produced, which requires a
more general, non-perturbative treatment \citep{Fl_2006a,
Fl_Biet_2007} in which the DSR spectrum has the form:
\begin{equation}
\label{I_all_02} I_{\omega}=\frac{8Q^2q(\omega)}{3 \pi c} \gamma^2
\left(1+ \frac{\omega\pe^2 \gamma^2}{\omega^2} \right)^{-1} \Phi(s),
\end{equation}
where $\Phi(s)=24 s^2   \int_{0}^{\infty} d t \exp(-2s t) \sin (2st)
\left[ \coth t-\frac{1}{t} \right]$ is the Migdal's function,  which
depends on a single parameter $s=\frac{1}{8\gamma^2} \left(
\frac{\omega}{q(\omega)} \right)^{1/2} \left( 1+ \frac{\omega\pe^2
\gamma^2}{\omega^2} \right)$; the scattering rate $q(\omega)$ has
been defined by Eq.~(\ref{Intensity_q_pereturb}).

Therefore, the DSR intensity depends on the following (microscopic)
parameters: the electron plasma frequency $\omega\pe$, the (defined
by the random field) gyrofrequency $\omega\st$, the frequency
$\omega_0=q_{\min} c$ specified by the main scale
$L_{\max}=2\pi/q_{\min}$ of the random field spectrum, the spectral
index $\nu$ of the random field, the Lorenz-factor $\gamma$ of the
emitting relativistic electrons, and also on number (and spectral
distribution) of the emitting electrons.

\subsection{Links between microscopic and macroscopic parameters of GRBs}

One of the difficulties in direct modeling of the GRB spectra is
that they depend on microscopic parameters, which are basically
unknown. However, we can link many of them with the macroscopic and
phenomenological parameters of the GRB source introduced in
\S~\ref{S_basic_parms}, \citep[sf, e.g.,][]{Kumar_McMahon_2008}.

The electron plasma frequency is specified by the electron number
density $n_e$
\begin{equation} \label{omega_p02}
\omega\pe=\sqrt{\frac{4 \pi n_e e^2}{m_e}}.
\end{equation}
Here we have to substitute the number density (\ref{n_01}) found
from the estimates of the total number of particles
(\ref{N_chislo_chastits}) and the source volume (\ref{volume_01}),
then
\begin{equation} \label{omega_p03}
\omega\pe=\sqrt{\frac{4 \pi N e^2}{m_e V'}}=\sqrt{\frac{4\pi e^2
}{m_e V'}\frac{E}{\Gamma m_p c^2}}= \sqrt{\frac{4\pi e^2
}{m_e}\frac{3}{4 \pi \Omega_{ang} R^3} \frac{E}{\Gamma m_p c^2}}=
 $$$$
\sqrt{\frac{3 e^2 E}{ m_e \Omega_{ang} (2c\delta t \Gamma)^3 \Gamma
m_p c^2}}=\sqrt{\frac{3 e^2 E}{ 8 m_e \Omega_{ang} \delta t^3
\Gamma^4 m_p c^5}}
\end{equation}
where  $e$ and $m_e$ are the electron charge and mass, $m_p$ is the
proton mass, 
$\delta t$ is the observed
time of the emission variability, $\Omega_{ang}$ is the jet opening
angle, and $E$ is the bulk kinetic energy of the shell. Therefore,
the plasma frequency is expressed through a number of values, which
can either be directly observed (like $\delta t$) or estimated from
the observations (like $E$, $\Gamma$, and $\Omega_{ang}$).

Similarly, the stochastic gyrofrequency
\begin{equation}
\label{omega_st_B02} \omega\st^2=\frac{e^2 \langle
B\st^2\rangle}{m_{e}^2 c^2}
\end{equation}
is defined by the random magnetic field value, and so can be
expressed via the phenomenological parameter $\epsilon_B$ describing
the energy content of the magnetic field:

\begin{equation}
\label{omega_st_B} \omega\st=\sqrt{\frac{8 \pi e^2 \epsilon_B
E}{m_{e}^2 c^2 V'}}=\sqrt{\frac{8 \pi e^2 \epsilon_B E}{m_{e}^2 c^2
\frac{4 \pi}{3} \Omega_{ang} (2c\delta t \Gamma)^3}}= \sqrt{\frac{3
e^2 \epsilon_B E} {4 m_{e}^2 c^5 \Gamma^3 \delta t^3}}.
\end{equation}

The third parameter of the DSR spectrum having the dimension of
frequency is $\omega_0=q_{\min}c=2\pi c/L_{\max}$, which is
determined by the main scale of the random magnetic field
distribution. To parameterize this unconstrained value we introduce
a dimensionless model parameter $\Lambda=\omega_0/\omega\st$ so that
\begin{equation}
\label{omega_0_1} \omega_{0}=\Lambda \cdot \omega\st.
\end{equation}
Then, $\Lambda \gg 1$ corresponds to a weak random field, while
$\Lambda \ll 1$ to a strong random field.

Finally, the distribution of the magnetic energy over the spatial
scales is determined by the spectral index $\nu$. We have no
reliable constraint for this value from the GRB observation,
however, using analogy with other astrophysical and laboratory
cases, we adopt $1< \nu <2$.

Consider now parameters characterizing emitting relativistic
electrons. The typical Lorenz-factor $\gamma$  of electrons can be
expressed from Eq.~(\ref{ee}) via parameters $\epsilon_e$, $\xi_e$,
and $\Gamma$:

\begin{equation}
\label{gamma_e01} \gamma=\frac{\epsilon_e \Gamma m_p}{\xi_e m_e}.
\end{equation}
Therefore, a single DSR spectrum is fully determined by specifying
the following set of ten involved parameters: $E$, $\Gamma$,
$\epsilon_e$, $\epsilon_B$, $\Lambda$, $\xi_e$, $\delta t$,
$\Omega_{ang}$, $\nu$, and $z$. We turn now to specifying the
corresponding parameter ranges.

\subsection{Modeling strategy}

For the sake of further modeling of the GRB spectra within the DSR
emission model we note that the spectral parameters $\alpha$,
$\beta$, and $E_{\rm break}$ of the Band fitting function display
single-mode distributions. In terms of statistical properties of the
GRB sources this implies that the GRB physical parameters are taken
from the same parent distributions.

Therefore, we will perform a kind of "global" statistical modeling
of the GRB spectra. To do so we produce a large number ($\sim$5000)
of individual DSR spectra, which differ from each other because
different combination of input parameters is selected for each
individual spectrum. Specifically, the ten independent parameters
are randomly selected from the corresponding parent distributions,
while the dependent (derived) parameters are then calculated as
described in the previous section.

The independent parameters, where known, are taken based on
observations available and the standard fireball model. For example,
we adopt the total shell energy $E\sim10^{51}$~erg, the jet opening
angle $\Omega_{ang}\sim10^{-3}$~ster, and $\Gamma\sim 300~(> 100)$
to have normal distributions around the mean values, while the time
variability scale $\delta t$ to have a log-normal distribution with
the mean $\langle \delta t \rangle=10^0$~s and standard deviation
$\sigma(\delta t)=10^{0.3}$~s as observed \citep{Nakar_Piran_2002}.
Parameters $\epsilon_e$ and $\epsilon_B$ are poorly constrained by
observations; we kept them much less than unity in all cases, while
consider a number of distributions for the spectral index $\nu$,
which is also unknown parameter of the model. To reduce the number
of free model parameters, we kept some parameters constant, which
have only minor effect on the DSR spectrum shape; namely, we adopted
$z=2$ and $\xi_e=1$. 

Then, we made a simplifying assumption about the energy spectrum of
the shock-accelerated electrons. Although this energy distribution
is likely to be a broad one (for example, a power-law $n_e(E)\propto
E^{-p}$), which can easily be taken into account, we adopt here a
monoenergetic electron distribution. Indeed, adding the power-law
energy distribution to the model would yield some spectrum regimes
common for DSR and other competing emission processes including
standard synchrotron radiation. We, however, want to evaluate the
capability of the DSR mechanism itself to reproduce the Band
function parameter distributions. If we succeed to do this with the
monoenergetic distribution, then the power-law distribution will
also be an acceptable one, since this will only increase (but not
decrease) the variety of the produced radiation spectra; we return
to this point later.

Finally, parameter $\Lambda$ is unknown. To evaluate the range of
this parameter the most consistent with observed distributions of
the GRB spectra, we vary it in a broad limits in our modeling from
run to run, while keep a constant within each run.

\section{Modeling results}

\subsection{Dependences of the DSR spectrum on the input parameters}
\label{S_DSR_spectra}

Before turning to the statistical modeling described, it is
worthwhile to consider how the DSR spectra depend on the independent
input parameters. These dependences are not self-evident, because
each input parameter (e.g., the bulk Lorenz-factor or bulk energy of
the shell) affect a few microscopic parameters, which, in their
turn, specify the DSR spectrum produced.

To do so, let us select the  following set of "standard" parameters
$\Gamma=300$,  $\epsilon_e=10^{-2}$, $\epsilon_B=10^{-4}$, $\Lambda=
2$, $\xi_e=1$, $\delta t=1$~s, $\Omega_{ang}=10^{-3}$, $\nu=1.2$,
and $z=2$ and then consider how the DSR spectrum changes as one of
these parameters changes, while the others are kept the same.

\begin{figure} \hspace{-0.2in}\includegraphics[height=2.7in]{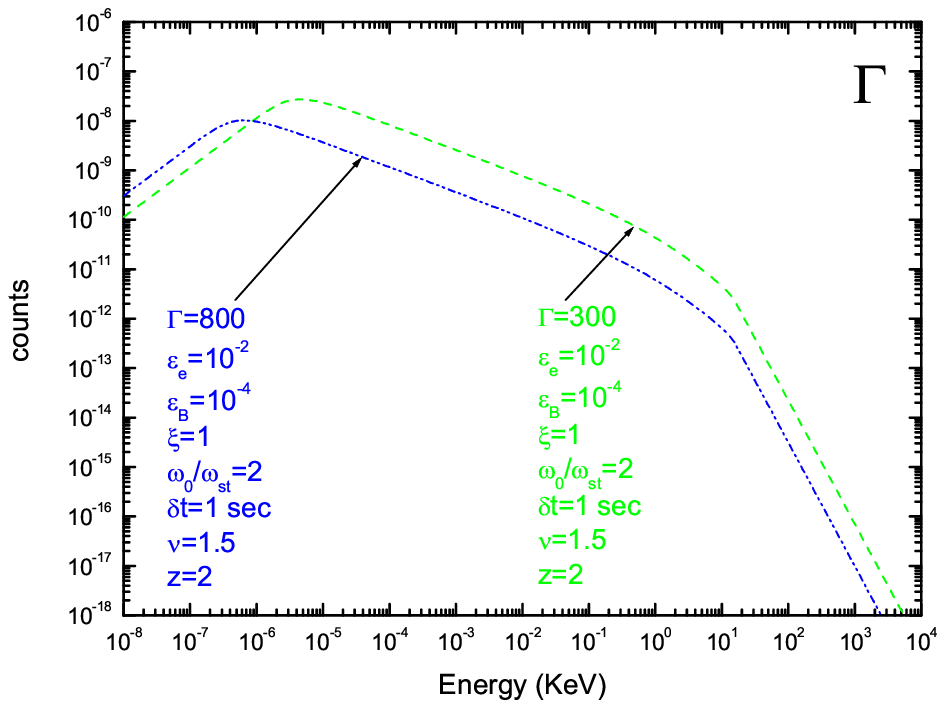}
 \hspace{-0.3in}\includegraphics[height=2.7in]{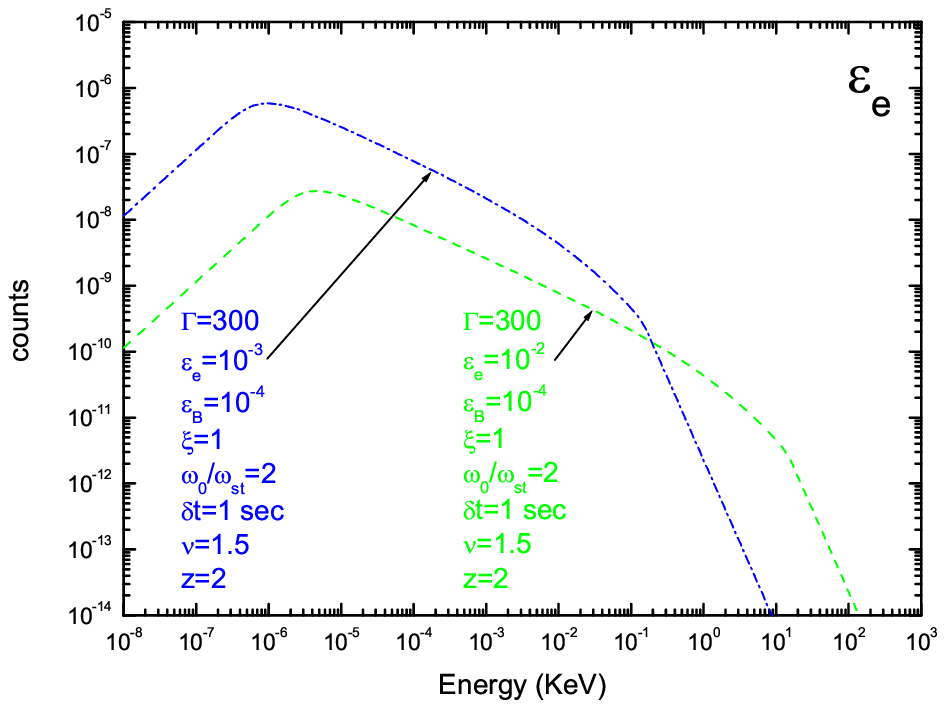}
 \caption{\small
Change in the DSR spectrum due to bulk Lorenz-factor $\Gamma$
change. \qquad\qquad\qquad\qquad\ }\label{razd_2_spectra_gamma}
 \caption{\small Change in the DSR spectrum due to $\epsilon_e$ change.}\label{razd_2_spectra_Ee}
\end{figure}

Figure~\ref{razd_2_spectra_gamma} shows that the DSR level decreases
and the spectrum moves towards \emph{lower} energies as the bulk
Lorenz-factor \emph{increases} in contrast to simple expectation
based on the bulk doppler shift only. The reason for such a behavior
is that the bulk Lorenz-factor (for the same other input parameters)
affects also other microscopic parameters, so the net effect of the
$\Gamma$ change in the fireball model is different from purely
kinematic Doppler effect. Increase of the relativistic electron
energy content $\epsilon_e$ shifts the whole spectrum towards higher
energies, Figure~\ref{razd_2_spectra_Ee}, since the main effect of
$\epsilon_e$ increase is the corresponding increase of the typical
Lorenz-factor $\gamma$ of the relativistic electrons.

\begin{figure} \hspace{-0.2in}\includegraphics[height=2.7in]{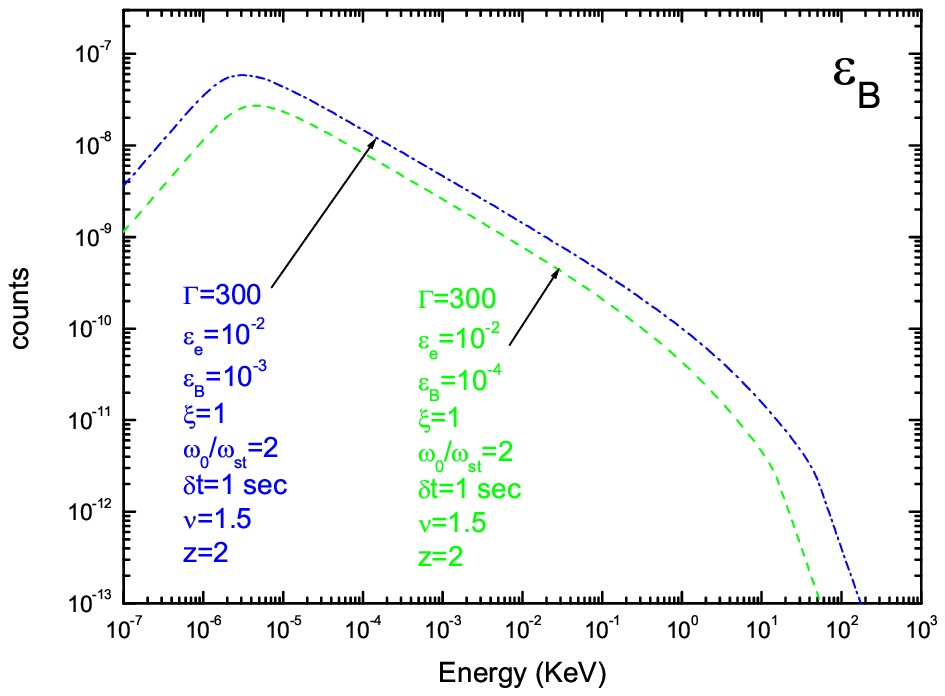}
 \hspace{-0.3in}\includegraphics[height=2.9in]{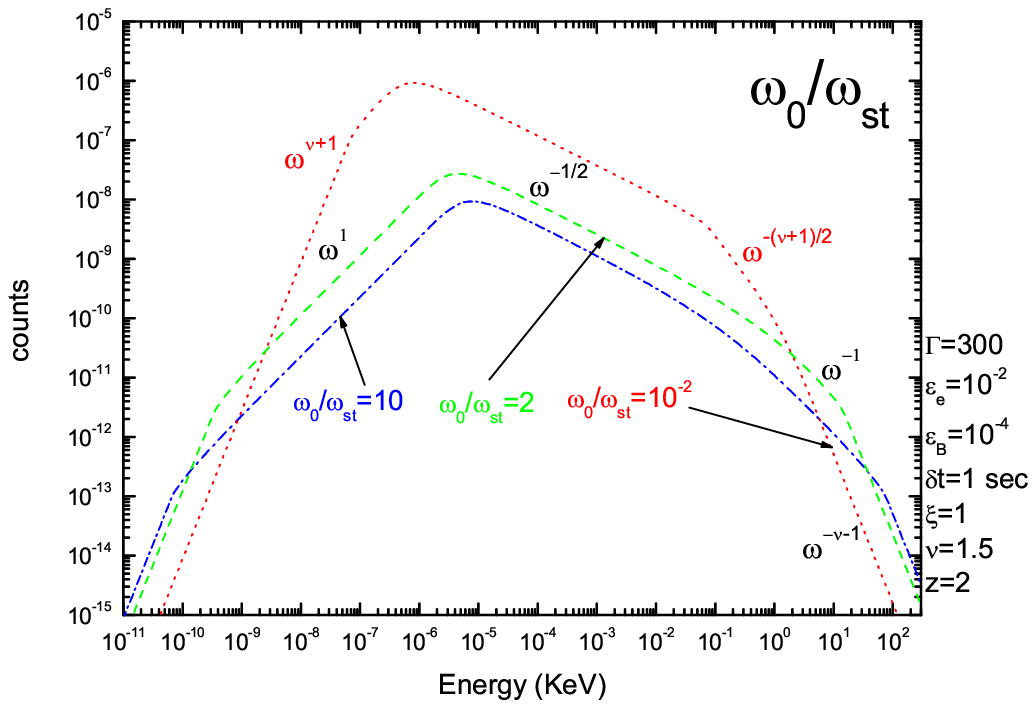}
\caption{\small Change in the DSR spectrum due to $\epsilon_B$
change }\label{razd_2_spectra_EB} \caption{\small Change in the DSR
spectrum due to $\Lambda=\omega_0/\omega\st$ change.
}\label{razd_2_spectra_om0omst}
\end{figure}

Figure~\ref{razd_2_spectra_EB} then displays that increase of the
magnetic energy density, $\epsilon_B$, results basically in the
upward shift of the whole spectrum and also to some shift of the
high-frequency part of the spectrum towards higher energies.
Parameter $\Lambda=\omega_0/\omega\st$ has a major effect on the DSR
spectrum Fig.~\ref{razd_2_spectra_om0omst} because it is the
parameter that controls, which DSR regime, perturbative or
non-perturbative, is in fact realized for a given parameter
combination, i.e., weak or strong random magnetic field is present
at the source. Accordingly, the DSR spectrum shape changes
significantly as $\Lambda$ changes, new spectrum asymptotes arise as
it decreases and non-perturbative DSR regime in a strong field
develops.

\begin{figure} \hspace{-0.2in}\includegraphics[height=2.7in]{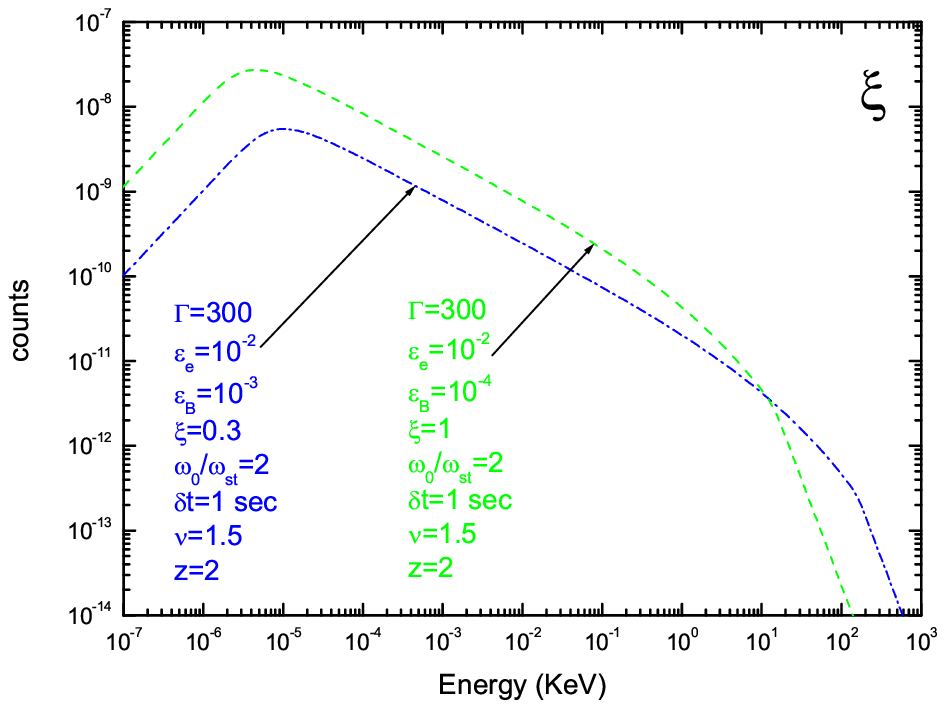}
 \hspace{-0.3in}\includegraphics[height=2.9in]{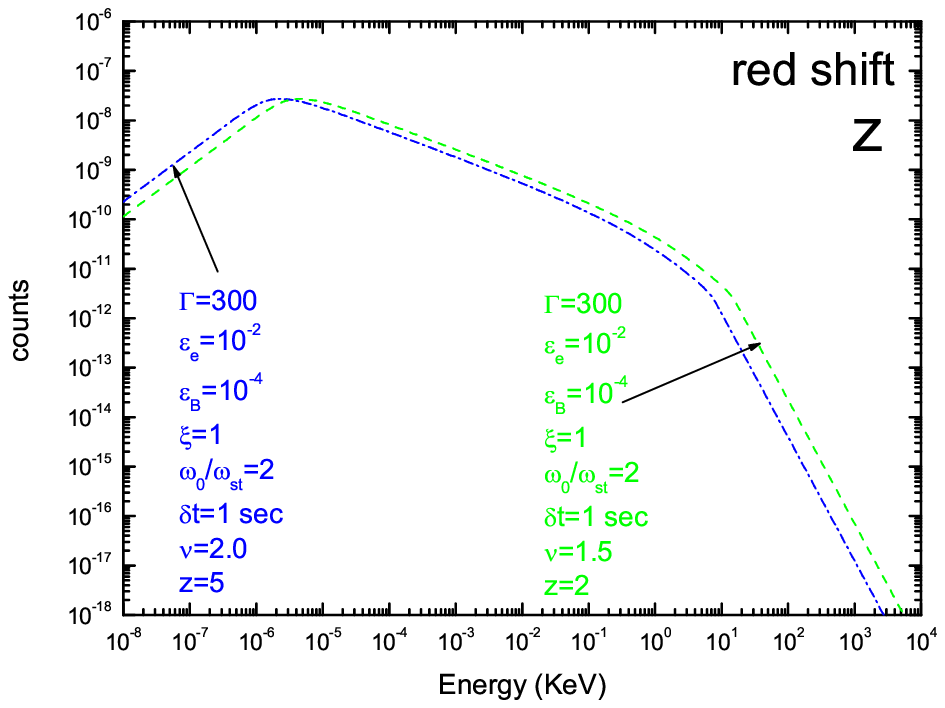}
 \caption{\small
Change in the DSR spectrum due to $\xi_e$ change.
}\label{razd_2_spectra_xi} \caption{\small Change in the DSR
spectrum due to $z$ change.}\label{razd_2_spectra_z}
\end{figure}

Decrease of the $\xi_e$ parameter results in a shift of the spectrum
towards higher energies, Fig.~\ref{razd_2_spectra_xi}. The reason
for that is the same as at $\epsilon_e$ increase: according to
Eq.~(\ref{gamma_e01}) both of them lead to $\gamma$ increase and so
to emission of correspondingly enhanced ($\propto \gamma^2$)
energies. Increase of the cosmological red shift parameter shifts
the whole spectrum towards lower energies,
Fig.~\ref{razd_2_spectra_z}, as can be expected from the kinematics
because no microscopic parameter depends on $z$ in this model.

\begin{figure} \hspace{-0.2in}\includegraphics[height=2.7in]{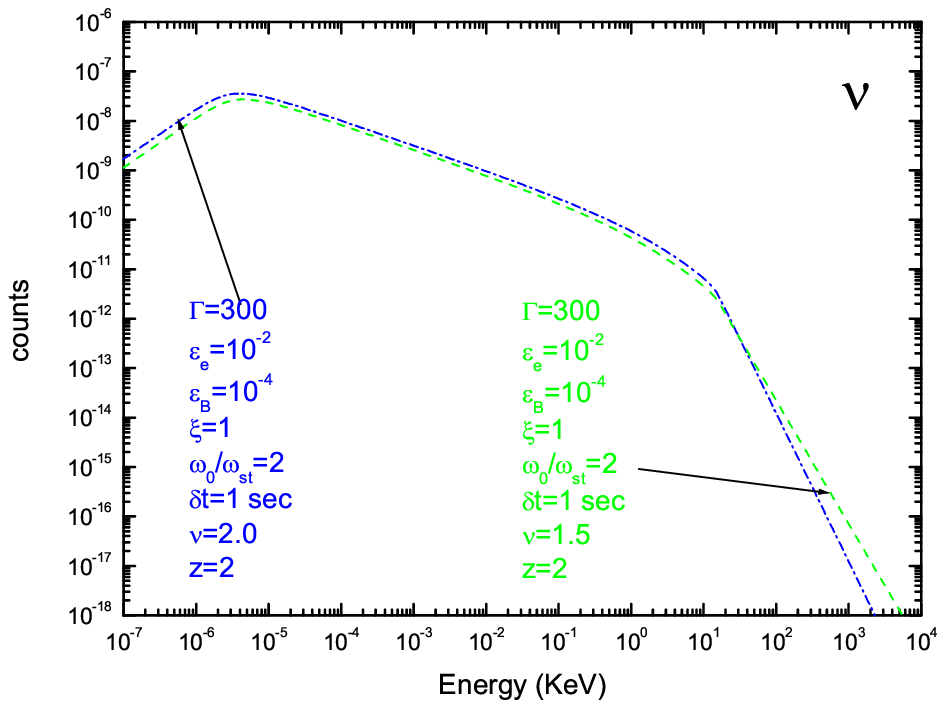}
 \hspace{-0.3in}\includegraphics[height=2.7in]{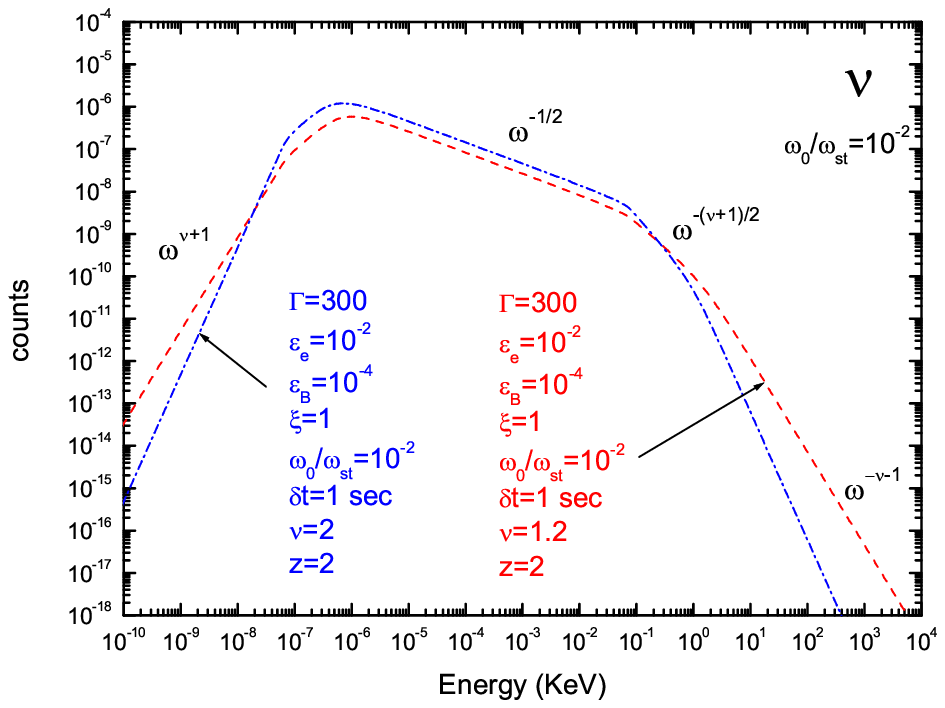}
 \caption{\small
 Change in the DSR spectrum due to $\nu$ change.}\label{razd_2_spectra_nu}
\end{figure}

Finally, the spectral index $\nu$ of the random field affects the
high-energy slope of the spectrum in the perturbative case, while
affects other spectrum asymptotes as well in a more general,
non-perturbative case, as can be seen from
Figure~\ref{razd_2_spectra_nu}. Now, when the important dependences
of the DSR spectrum on the input parameters are established, we are
in the position to perform the statistical modeling, and, to fine
tune the parameter ranges towards model distributions matching the
observed ones.

We start the modeling by adopting trial distributions for the
parameters we are going to vary within a single run. Specifically,
we initially adopt normal distributions for the involved parameters
with the following means and standard deviations:
$\langle\Gamma\rangle=300$, $\sigma_{\Gamma}=50$,
$\langle\epsilon_e\rangle=10^{-2}$, $\sigma_{\epsilon_e}=10^{-4}$,
$\langle\epsilon_B\rangle=10^{-4}$, $\sigma_{\epsilon_B}= 10^{-6}$,
$\langle\log_{10} \delta t\rangle=0$, $\sigma_{log_{10} \delta
t}=3\cdot10^{-1}$, $\langle\nu\rangle=1.2$, and
$\sigma_{\nu}=5\cdot10^{-2}$. Whenever possible, we will check if
other than normal distribution offers a better fit to the
observational data.

\subsection{Case I: Week random magnetic field; perturbative treatment applies}

\begin{figure}
 \centerline{
\includegraphics[height=3.0in]{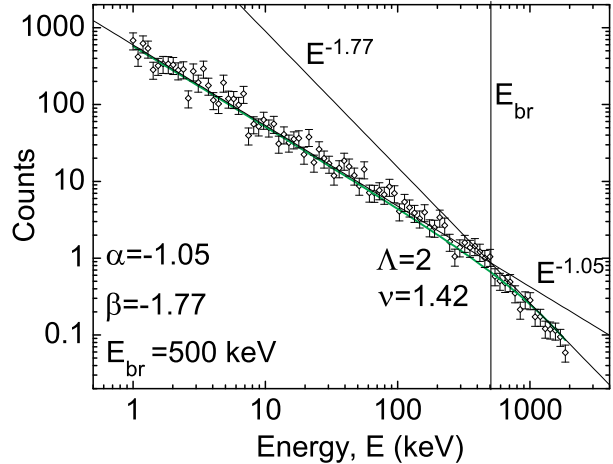}}
\caption{\small Example of the model noisy DSR spectrum (symbols,
$\Lambda=2$, $\nu=1.42$) and fitting Band spectrum (solid thick
green line). Thin solid lines display the low-energy and high-energy
power-law asymptotes of the fitting Band function and position of
the break energy; the Band parameters are shown in the left lower
part of the figure. Note, that the adopted $\nu$ value would imply
$\beta=-2.42$; however, relatively narrow region of the high-energy
part of the spectrum and also added noise result in flatter
asymptote with $\beta\approx -1.77$. } \label{fit_example}
\end{figure}

We begin with the simplest case when the DSR spectrum can be
described by a perturbative formulae derived in \cite{Fl_2006a},
which are widely used to evaluate the DSR in the perturbative
("jitter") regime \citep[see, e.g.,][]{Workman_etal_2008,
Mao_Wang_2007}. The condition for random magnetic field to be weak
is $\Lambda \equiv\omega_0/\omega\st\gg 1$; then, applicability of
the perturbative treatment requires additionally $\gamma <
\omega\pe\omega_0/\omega\st^2$ \citep{Fl_2006a}. Adopting a constant
value $\Lambda=10$ and taking randomly other involved parameters
from the parent normal distributions described in the previous
section, we generate around 5,000 individual DSR spectra, add noise
to them at the level of 25$\%$, and fit each of them to the
phenomenological Band function, see an example in
Figure~\ref{fit_example}. This yields distributions of the spectral
fitting parameters to be compared with the observed histograms of
the Band parameters.

\begin{figure}
\includegraphics[height=2.5in]{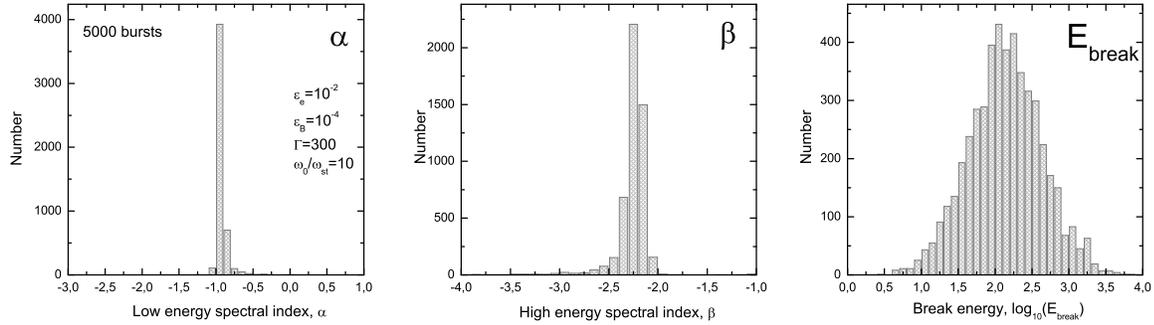}
\caption{\small Histograms of the Band parameters $\alpha$, $\beta$,
and $E\br$ obtained within the perturbative (jitter) DSR model
assuming the random magnetic field is
weak.}\label{razd_3_2_pereturb_hyst}
\end{figure}

Figure~\ref{razd_3_2_pereturb_hyst} displays the model histograms.
Remarkably, the model $E\br$ histogram is very similar to the
observed one, while  the $\alpha$ and $\beta$ histograms display the
peak values consistent with the observed ones ($-1$ and $-2.2$
respectively), although the widths of the model histograms are much
smaller than of the observed ones. This inconsistency can be related
to (i) use of the simplified perturbative treatment, (ii)
non-optimal parameter range, or (iii) fundamental shortage of the
adopted DSR model. We, thus, address these issues by applying full
non-perturbative DSR treatment and exploring more complete range of
the involved parameters. The remaining residuals between the model
and observations will then be critically discussed within
simplifications and limitations of the model used.

\subsection{Case II: Week random magnetic field; non-perturbative treatment}
\label{S_weak_Npert}

As is known \citep{Topt_Fl_1987, Fl_2006a} non-perturbative
treatment may be required even for the case of relatively weak
random field if the energy of radiating electron is large; it
results in the $\propto \omega^{-1/2}$ asymptote, see, e.g.,
Figure~\ref{razd_2_spectra_om0omst}. It is quite clear that an
immediate outcome of this non-perturbative asymptote is appearance
of $\alpha$ values around $-1/2$, so the range of $\alpha$ values
between $-1$ and $-1/2$ will be filled. The distribution of
individual $\alpha$ values in this range will change depending on
the adopted $\Lambda=\omega_0/\omega\st$ value.

\begin{figure}
\includegraphics[height=2.5in]{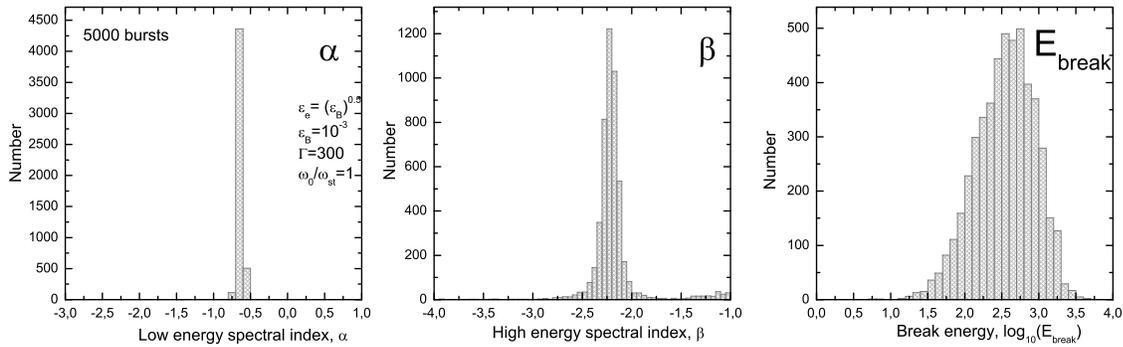}
\caption{\small  Histograms of the Band parameters $\alpha$,
$\beta$, and $E\br$ obtained within the non-perturbative DSR model
assuming the random magnetic field is weak,
$\Lambda=1$.}\label{razd_3_3_hyst_gauss}
\end{figure}

Figure~\ref{razd_3_3_hyst_gauss} displays an example of the Band
parameter distributions obtained within the non-perturbative
treatment for the case $\Lambda=1$, where the energy content of the
magnetic field $\epsilon_B$  \citep[and, accordingly, $\epsilon_e$,
to keep $\epsilon_e=\sqrt{\epsilon_B}$, see, e.g.,][and references
therein]{Sironi_Spitkovsky_2009} is increased to
$\epsilon_B=10^{-3}$ to keep the break energy within the required
window of observations. We see that the peak of the $\alpha$
histogram shifts to the value $\approx-0.6$, related to the $\propto
\omega^{-1/2}$ asymptote. This histogram remains very narrow like in
the perturbative case, and, furthermore, its peak value does not
agree with the observed peak value any longer. Variation of
parameter $\Lambda$ within the range corresponding to the weak field
case, $\Lambda\ge1$ does not improve the situation: the distribution
remains narrow with the peak value between $-1$ and $-0.5$. We
conclude that the DSR model with the weak random magnetic field,
either perturbative or non-perturbative, cannot offer a consistent
fit to the observed $\alpha$ histogram, while the model $E\br$
histogram matches well the observed one. It is worthwhile to note
here that a model with a weak random field was also criticized from
another perspective \citep{Kumar_McMahon_2008} as it may imply an
unrealistically high level of inverse Compton emission. In addition,
\cite{Kirk_Reville_2010} argued that the weak-field case (needed to
mediate the jitter-like regime of DSR) seems to be in contradiction
with the required high efficiency of the particle acceleration at
the shocks and strong magnetic fluctuations are needed to
self-consistently accelerate electrons up to the gamma-ray producing
energies. Thus, we turn now to the case of the strong random
magnetic field.

\subsection{Case III: Strong random magnetic field}

Having the weak random field model (jitter regime) rejected, we turn
now to analysis of the strong random field case, $\Lambda\le1$. As
has been explained in \S~\ref{S_DSR_spectra}, new asymptotes,
including $\propto\omega^{-(\nu+1)/2}$, arise in this case, which
can yield broader $\alpha$ distribution if this new asymptote comes
into play. Therefore, the model $\alpha$ distribution will depend on
adopted $\nu$ distribution, which is in fact unconstrained by the
observations. However, we can take advantage of the fact that in our
model the $\beta$ distribution is straightforwardly determined by
the $\nu$ distribution. Indeed, because $\beta=-\nu-1$, we can
simply derive the required $\nu$ distribution from the observed
$\beta$ distribution.

\begin{figure}
\includegraphics[height=2.5in]{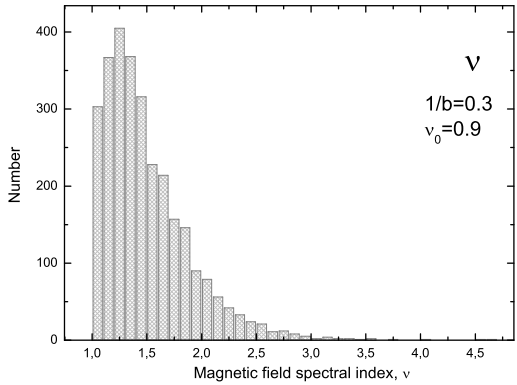}
\caption{\small Example of Gamma distribution of the spectral index
$\nu$ used in the modeling.}\label{razd_3_3_nu_hyst0903}
\end{figure}

The observed $\beta$ histogram reaches the peak at around $-2.2$ and
has asymmetric skew shape with a longer tail towards smaller values,
which apparently cannot be described by a symmetric normal
distribution adopted above. Thus, we have to adopt another
reasonably simple distribution roughly matching the observed one.
Specifically, we find that the Gamma distribution
\begin{equation}
\label{Gamma_distr_01} F(\nu)=(\nu-\nu_0)^{R-1}\cdot
\frac{e^{-(\nu-\nu_0)/b}}{b^R\Gamma(R)},
\end{equation}
with $R=2$, where $\Gamma(R)$ is the Euler Gamma function, is well
suited for our modeling. This distribution has a peak at
$\nu=\nu_0+b$; we expect to obtain correct $\beta$ histogram for
$\nu_0=0.9-1$ and $b=0.3-0.2$, see
Figure~\ref{razd_3_3_nu_hyst0903}.

\begin{figure}
\includegraphics[height=2.5in]{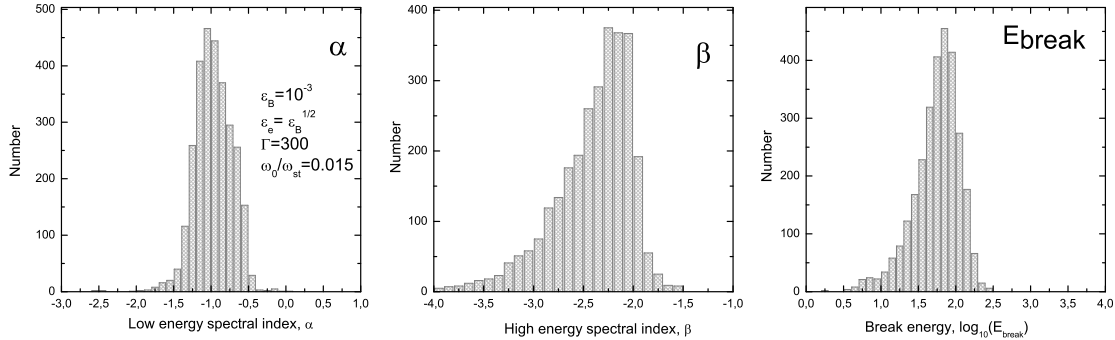}
\caption{\small Example of the model Band parameter ($\alpha$,
$\beta$, and $E\br$) distributions obtained within the DSR model
with strong random magnetic field.}\label{razd_3_3_hyst0903}
\end{figure}

Other than the $\nu$ distribution adopted to obey Gamma distribution
with the specified parameters, all other steps of the modeling are
the same as before. We performed many runs changing the $\Lambda$
value and also varying other involved parameters within the adopted
limits and found that typically the $\alpha$ distribution is much
broader than in the case of the weak field considered in two
previous sections in a general agreement with observations; the peak
value of the $\alpha$ histogram varies between $-1.1$ and $-0.8$
when $\Lambda<1$. Figure~\ref{razd_3_3_hyst0903} displays an example
of the model Band spectral parameters obtained for the strong random
field case.

The model results are in a remarkable agreement with the
observations. Indeed, the $\alpha$ histogram is a symmetric one, it
displays a peak at the right place,  $\alpha=-1$, and its bandwidth
is comparable to that of the observed histogram. The $\beta$
histogram almost repeats the observed one, displaying the correct
asymmetric shape and the peak at the right place,  $\beta=-2.2$. The
$E\br$ histogram agrees with the observed one rather well: it has
correct shape and bandwidth, although the peak value is less than
the observed value by the factor around 2. This discrepancy is,
however, inessential: as we have seen in \S~\ref{S_weak_Npert}, the
position of the break energy can easily be adjusted by a small
change of the magnetic energy content $\epsilon_B$ and corresponding
change of $\epsilon_e$. Moreover, considering broader range of
$\xi_e$ variation (recall, we adopted a constant value of $\xi_e =1$
in our modeling) can also easily change the characteristic break
energy by a factor of 2 or more. Thus, we can conclude that the
simplified DSR model adopted in this section is intrinsically
capable of reproducing the Band parameter distributions compatible
with the observed ones.

\subsection{Cross-correlations between the model Band parameters}

In addition to the Band parameter distributions themselves, it is
worthwhile to address a question if our model reproduces the
cross-correlations between the Band parameters correctly. This task
can easily be solved using the variety of the model spectra produced
in each model run.

\begin{figure}
\includegraphics[height=2.05in]{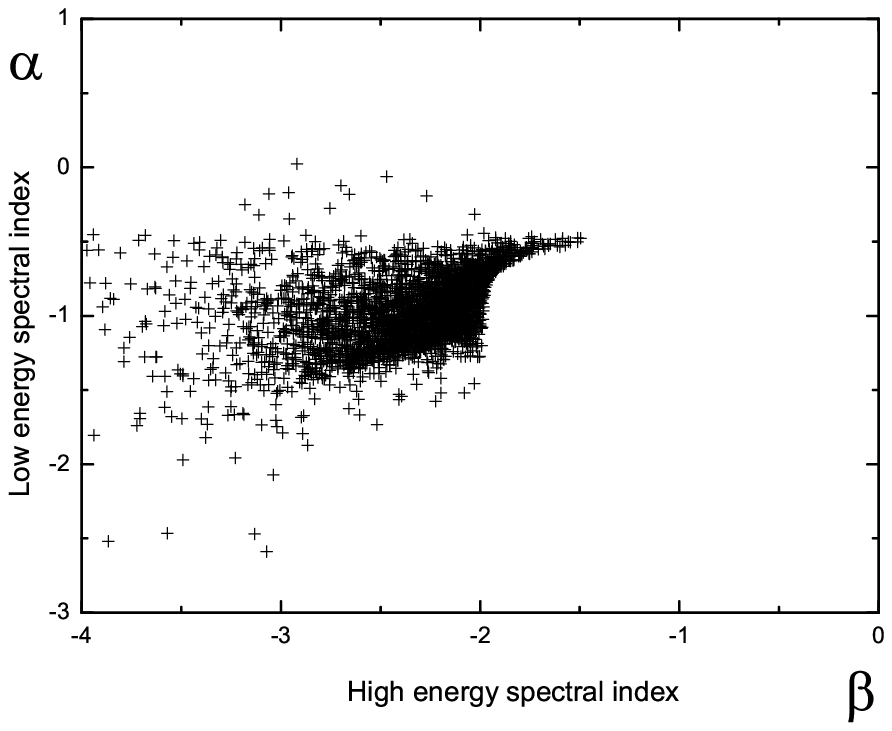}\hspace{-1cm}
\includegraphics[height=2.05in]{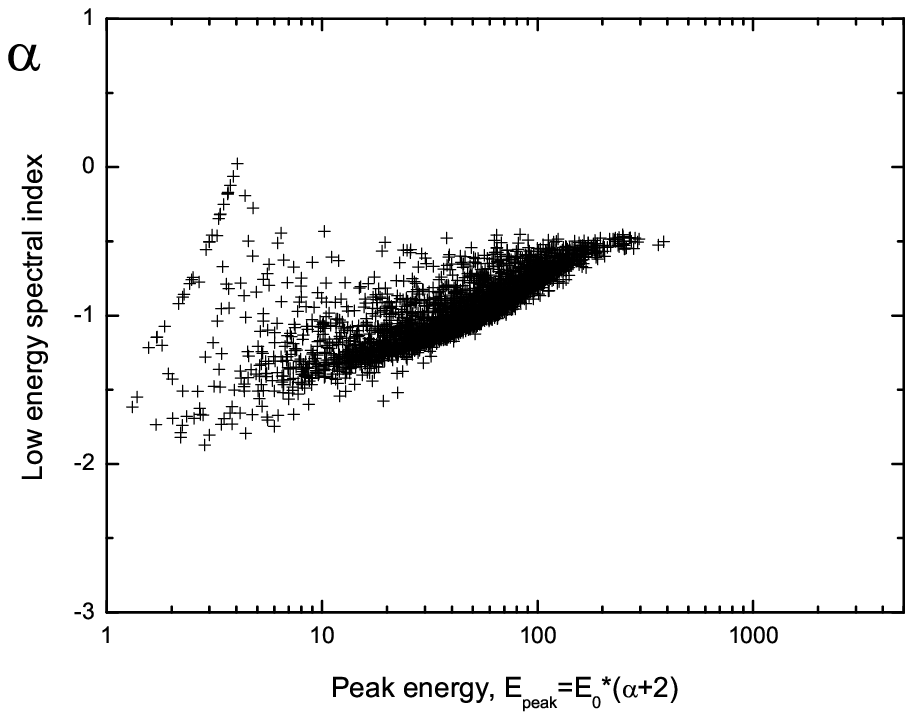}\hspace{-1cm}
\includegraphics[height=2.05in]{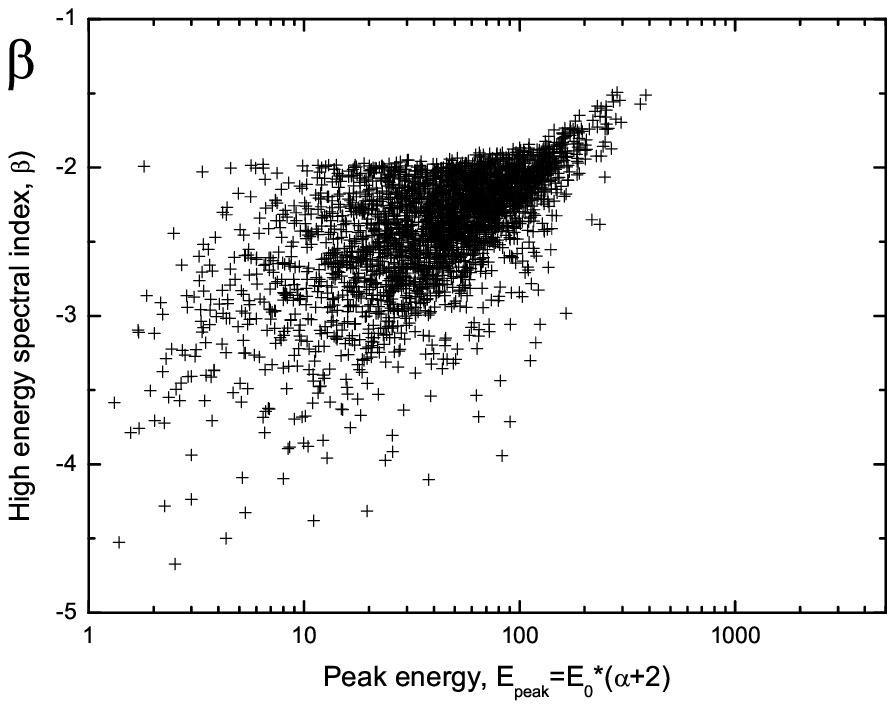}
\caption{\small Cross-correlation of the model Bend spectral
parameters $\alpha$ and $\beta$.} \label{razd_3_7_correl_a_b}
\caption{\small Cross-correlation of the model Bend spectral index
$\alpha$ and $E_{peak}=(\alpha+2)E_0$.} \label{razd_3_7_correl_a_E}
\caption{\small Cross-correlation of the model Bend spectral index
$\beta$ and $E_{peak}=(\alpha+2)E_0$.} \label{razd_3_7_correl_b_E}
\end{figure}
%

Figures~\ref{razd_3_7_correl_a_b}--\ref{razd_3_7_correl_b_E} display
these cross-correlations to be compared with Figure~31 from
\citet{Kaneko_etal_2006}. Like in the observation, the spectral
indices $\alpha$ and $\beta$ are not highly correlated, although in
the model plot the region of $-0.5<\alpha<0$ is underpopulated
compared with the observed plot \citep{Kaneko_etal_2006}. Two other
plots are in remarkable agreement with the observed
cross-correlation plots, presented in \cite{Kaneko_etal_2006}. We
conclude that the developed model is naturally capable of
reproducing the cross-correlation plots in addition to the
histograms themselves, which is a remarkable success of the
non-perturbative DSR model in the presence of strong random magnetic
field.

\section{Discussion}

Our modeling shows that we can get an overall 
agreement between the model and observed histograms of the Band GRB
spectral parameters within the non-perturbative DSR model with
strong random magnetic field when $\Lambda \equiv
\omega_0/\omega\st\approx 0.015$. We note that in our model the
$\Lambda$ parameter is not derived from a microscopic treatment of
the shock interactions, rather it is a free model parameter adjusted
for the model histogram to resemble the observed ones. Let us
consider if the obtained $\Lambda$ value has any sense versus
current models of magnetic field generation at the shock fronts. To
do so we recall that $\omega_0$ is defined by the correlation length
$L_{\max}$ of the random magnetic field, $\omega_0=2\pi c/L_{\max}$.
If the magnetic field is produced by a two-stream (filamentation or
Weibel) instability, the correlation length $L_{\max}$ is expected
to be about ten plasma skin scales, $l_{sc}\sim 10 c/\omega\pe$
\citep[e.g.,][]{Sironi_Spitkovsky_2009}. 
Since the frequencies $\omega\pe$ and
$\omega_0$ are independent input parameters in the modeling, we can
check if the correlation length is indeed of the order of the skin
scale, by inspecting the actual distribution of the
$\omega_0/\omega\pe$ ratio as it appears in the best-fit model case.

\begin{figure} 
\hspace{-0.4in}\includegraphics[height=3.2in]{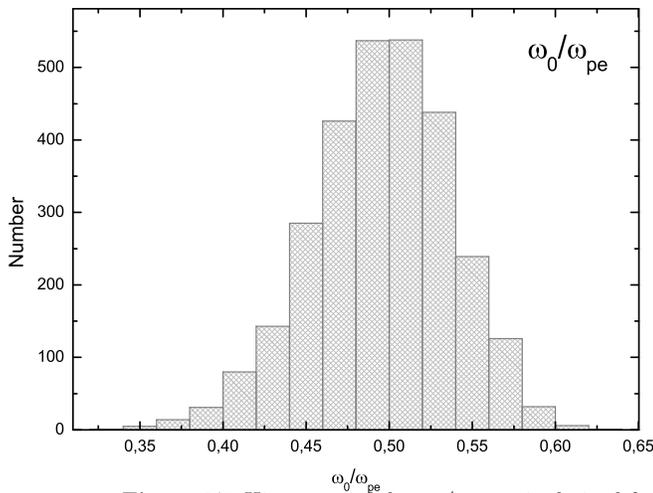}
\vspace{-0.4in} \caption{\small Histogram of the
$\omega_0/\omega\pe$ ratio derived from the DSR model with the
strong random field with $\Lambda=0.015$.}\label{razd_5_hysto_w0_wp}
\end{figure}

Figure~\ref{razd_5_hysto_w0_wp} displays a histogram of the
$\omega_0/\omega\pe$ ratio obtained for the model with $\Lambda =
0.015$. The distribution has a symmetric bell shape with the peak
about 0.5, thus, 
$L_{\max}\approx 4\pi l_{sc}$ and the required random field
correlation length is indeed of the order of ten plasma skin scales,
which agrees with the idea of the random field generation by a
two-steam instability in the internal shock interactions.

We must note that although the model histograms and the
cross-correlation plots look similar to the observed ones, there are
some residuals between them. For example, the model $\alpha$
histogram is somewhat narrower than the observed one with a deficit
of values $-2<\alpha<-1.5$ and $-0.5<\alpha<0$. The first interval
can possibly be filled if one considers a nonuniformity of the
emission source, which can naturally broaden the emission spectrum
leading eventually to smaller $\alpha$ values. The second interval
requires some additional physics, not included in our simplified
modeling, to be taken into account, for example, random electric
fields, turbulence anisotropy, or specific source geometry/viewing
angle combination.

On the other hand, given the number of simplifying assumptions
adopted for the modeling, we can conclude that the obtained
agreement between the model and observations is remarkably good. Let
us briefly remind and discuss those simplifications.

\begin{enumerate}
    \item We considered random magnetic fields only. It is known,
however, that diffusive radiation in electrostatic Langmuir waves
results in a harder spectrum, e.g. $I_\omega\propto \omega^1$,
\citep{Fl_Topt_2007a, Fl_Topt_2007b}, thus, the presence of these
Langmuir waves could compensate the deficit of the $-0.5<\alpha<0$
values.
 \item A monoenergetic spectrum of accelerated electrons was
adopted. Although typically the distribution of $\beta$ parameter is
ascribed to a parent distribution of the spectral index $p$ of the
electron distribution over energy, $N(E)\propto E^p$, our modeling
shows that the right distribution of the $\beta$ index can easily be
obtained even for a monoenergetic electron spectrum $N(E)\propto
\delta(E-E_*)$. In fact, for power-law energy distributions almost
the same results hold for $p> 2\nu+1$.
However, if a power-law range with $p< 2\nu+1$ is present in the
electron energy spectrum, this is not in a contradiction with the
model, although this does add more flexibility to formation of the
$\beta$ distribution, which can further broaden it towards even
better similarity to the observed histogram.
 \item A number of the source parameters (e.g., $z$ and $\xi_e$) were
adopted to be the same for all the sources; in fact, adopting more
realistic distributions can broaden the obtained distributions of
the Band spectral parameters and also affect the best-fit parameter
space, so the accuracy of the obtained source parameters is at best
to an order of magnitude.
 \item Random magnetic fields accounted by the model were adopted to
be statistically uniform and isotropic. Inclusion of the turbulence
anisotropy can  affect the radiation spectra and so modify the Band
parameter distributions.
 \item Having adopted both turbulence and electrons are
isotropically distributed, we did not consider any source
geometry/viewing angle effect. For the case of anisotropic
distributions such effects can also come into play.
 \item And finally, we did not explicitly consider the source
evolution, although both particle distribution and magnetic field
can evolve in time resulting in a GRB spectral evolution.
\end{enumerate}

The latter three simplifications have in fact been addressed
\citep[e.g.,][and references therin]{Workman_etal_2008} in a number
of studies, however, a purely perturbative (jitter) weak-field
regime of the DSR was adopted to calculate the radiation spectra.
This weak-field regime was shown to rise severe problems in
application to the GRB prompt emission \citep{Kumar_McMahon_2008,
Kirk_Reville_2010}. In addition, as has been demonstrated above in
this paper, this jitter regime of the DSR is inconsistent with the
observed Band parameter distributions, while the strong-field regime
is in fact needed. Since the strong-field regime requires a much
more sophisticated fully non-perurbative treatment, those previous
studies cannot be straightforwardly applied to this case, and so the
required generalization 
must be specifically performed from scratch 
within the non-perturbative treatment \citep{Topt_Fl_1987, Fl_2006a,
Fl_Biet_2007}.

Although all these effects are potentially important and must
eventually  be taken into account in building a more comprehensive
model, we conclude that even the simplified DSR model considered
here is naturally capable of reproducing main characteristic
properties of the Band parameter distributions and
cross-correlations between them. The ranges of the parameters needed
for the model to most closely reproduce the observed histograms
agree well with standard fireball model parameters.

\section*{Acknowledgments}

This work was supported in part by the Russian Foundation for Basic
Research, grants No. 08-02-92228, 09-02-00226, 09-02-00624. We have
made use of NASA's Astrophysics Data System Abstract Service.


\bibliographystyle{mn2e} 
\bibliography{DSR_PWNs,grb,fleishman}

\begin{thebibliography}{}

\bibitem[\protect\citeauthoryear{{Band}, {Matteson}, {Ford}, {Schaefer},
  {Palmer}, {Teegarden}, {Cline}, {Briggs}, {Paciesas}, {Pendleton}, {Fishman},
  {Kouveliotou}, {Meegan}, {Wilson} \& {Lestrade}}{{Band}
  et~al.}{1993}]{Band1993}
{Band} D.,  {Matteson} J.,  {Ford} L.,  {Schaefer} B.,  {Palmer} D.,
  {Teegarden} B.,  {Cline} T.,  {Briggs} M.,  {Paciesas} W.,  {Pendleton} G.,
  {Fishman} G.,  {Kouveliotou} C.,  {Meegan} C.,  {Wilson} R.,    {Lestrade}
  P.,  1993, \apj, 413, 281

\bibitem[\protect\citeauthoryear{{Baring} \& {Braby}}{{Baring} \&
  {Braby}}{2004}]{Baring_Braby_2004}
{Baring} M.~G.,  {Braby} M.~L.,  2004, \apj, 613, 460

\bibitem[\protect\citeauthoryear{{Bret} \& {Dieckmann}}{{Bret} \&
  {Dieckmann}}{2008}]{Bret_Dieckmann_2008}
{Bret} A.,  {Dieckmann} M.~E.,  2008, Physics of Plasmas, 15, 062102

\bibitem[\protect\citeauthoryear{{Bret}, {Firpo} \& {Deutsch}}{{Bret}
  et~al.}{2004}]{Bret_2004}
{Bret} A.,  {Firpo} M.,    {Deutsch} C.,  2004, \pre, 70, 046401

\bibitem[\protect\citeauthoryear{{Bret}, {Firpo} \& {Deutsch}}{{Bret}
  et~al.}{2005}]{Bret_2005}
{Bret} A.,  {Firpo} M.,    {Deutsch} C.,  2005, Physical Review Letters, 94,
  115002

\bibitem[\protect\citeauthoryear{{Bykov} \& {Meszaros}}{{Bykov} \&
  {Meszaros}}{1996}]{Bykov_Meszaros_1996}
{Bykov} A.~M.,  {Meszaros} P.,  1996, \apjl, 461, L37+

\bibitem[\protect\citeauthoryear{{Dieckmann} \& {Bret}}{{Dieckmann} \&
  {Bret}}{2010}]{Dieckmann_Bret_2010}
{Dieckmann} M.~E.,  {Bret} A.,  2010, \physscr, 81, 015502

\bibitem[\protect\citeauthoryear{{Fleishman}}{{Fleishman}}{2006}]{Fl_2006a}
{Fleishman} G.~D.,  2006, \apj, 638, 348

\bibitem[\protect\citeauthoryear{{Fleishman} \& {Bietenholz}}{{Fleishman} \&
  {Bietenholz}}{2007}]{Fl_Biet_2007}
{Fleishman} G.~D.,  {Bietenholz} M.~F.,  2007, \mnras, 376, 625

\bibitem[\protect\citeauthoryear{{Fleishman} \& {Toptygin}}{{Fleishman} \&
  {Toptygin}}{2007a}]{Fl_Topt_2007a}
{Fleishman} G.~D.,  {Toptygin} I.~N.,  2007a, \mnras, 381, 1473

\bibitem[\protect\citeauthoryear{{Fleishman} \& {Toptygin}}{{Fleishman} \&
  {Toptygin}}{2007b}]{Fl_Topt_2007b}
{Fleishman} G.~D.,  {Toptygin} I.~N.,  2007b, \pre, 76, 017401

\bibitem[\protect\citeauthoryear{{Frederiksen}, {Hededal}, {Haugb{\o}lle} \&
  {Nordlund}}{{Frederiksen} et~al.}{2004}]{Frederiksen_2004}
{Frederiksen} J.~T.,  {Hededal} C.~B.,  {Haugb{\o}lle} T.,    {Nordlund}
  {\AA}.,  2004, \apjl, 608, L13

\bibitem[\protect\citeauthoryear{{Granot}, {Fermi LAT} \& {GBM
  collaborations}}{{Granot} et~al.}{2009}]{Granot_etal_2009}
{Granot} J.,  {Fermi LAT} f.~t.,    {GBM collaborations} 2009, ArXiv e-prints

\bibitem[\protect\citeauthoryear{{Hededal}}{{Hededal}}{2005}]{Hededel_PhD2005}
{Hededal} C.,  2005, PhD thesis, , Niels Bohr Institute

\bibitem[\protect\citeauthoryear{{Hededal} \& {Nishikawa}}{{Hededal} \&
  {Nishikawa}}{2005}]{Hededal_Nishikawa_2005}
{Hededal} C.~B.,  {Nishikawa} K.-I.,  2005, \apjl, 623, L89

\bibitem[\protect\citeauthoryear{{Jaroschek}, {Lesch} \&
  {Treumann}}{{Jaroschek} et~al.}{2004}]{Jaroschek_2004}
{Jaroschek} C.~H.,  {Lesch} H.,    {Treumann} R.~A.,  2004, \apj, 616, 1065

\bibitem[\protect\citeauthoryear{{Kaneko}, {Preece}, {Briggs}, {Paciesas},
  {Meegan} \& {Band}}{{Kaneko} et~al.}{2006}]{Kaneko_etal_2006}
{Kaneko} Y.,  {Preece} R.~D.,  {Briggs} M.~S.,  {Paciesas} W.~S.,  {Meegan}
  C.~A.,    {Band} D.~L.,  2006, \apjs, 166, 298

\bibitem[\protect\citeauthoryear{{Kazimura}, {Sakai}, {Neubert} \&
  {Bulanov}}{{Kazimura} et~al.}{1998}]{Kazimura_1998}
{Kazimura} Y.,  {Sakai} J.~I.,  {Neubert} T.,    {Bulanov} S.~V.,  1998, \apjl,
  498, L183

\bibitem[\protect\citeauthoryear{{Keshet}, {Katz}, {Spitkovsky} \&
  {Waxman}}{{Keshet} et~al.}{2008}]{Keshet_2008}
{Keshet} U.,  {Katz} B.,  {Spitkovsky} A.,    {Waxman} E.,  2008, in 37th
  COSPAR Scientific Assembly Vol.~37 of COSPAR, Plenary Meeting, {Evolution of
  magnetization in relativistic collisionless shocks}.
pp 1499--+

\bibitem[\protect\citeauthoryear{{Kirk} \& {Reville}}{{Kirk} \&
  {Reville}}{2010}]{Kirk_Reville_2010}
{Kirk} J.~G.,  {Reville} B.,  2010, \apjl, 710, L16

\bibitem[\protect\citeauthoryear{{Kumar} \& {McMahon}}{{Kumar} \&
  {McMahon}}{2008}]{Kumar_McMahon_2008}
{Kumar} P.,  {McMahon} E.,  2008, \mnras, 384, 33

\bibitem[\protect\citeauthoryear{{Mao} \& {Wang}}{{Mao} \&
  {Wang}}{2007}]{Mao_Wang_2007}
{Mao} J.,  {Wang} J.,  2007, \apjl, 669, L13

\bibitem[\protect\citeauthoryear{{Mazets}, {Aptekar}, {Frederiks},
  {Golenetskii}, {Il'Inskii}, {Palshin}, {Cline} \& {Butterworth}}{{Mazets}
  et~al.}{2004}]{Mazets_etal_2004}
{Mazets} E.~P.,  {Aptekar} R.~L.,  {Frederiks} D.~D.,  {Golenetskii} S.~V.,
  {Il'Inskii} V.~N.,  {Palshin} V.~D.,  {Cline} T.~L.,    {Butterworth} P.~S.,
  2004, in {M.~Feroci, F.~Frontera, N.~Masetti, \& L.~Piro} ed., Astronomical
  Society of the Pacific Conference Series Vol.~312 of Astronomical Society of
  the Pacific Conference Series, {Konus catalog of short GRBs}.
pp 102--+

\bibitem[\protect\citeauthoryear{{Medvedev} \& {Loeb}}{{Medvedev} \&
  {Loeb}}{1999}]{Medvedev_1999}
{Medvedev} M.~V.,  {Loeb} A.,  1999, \apj, 526, 697

\bibitem[\protect\citeauthoryear{{M{\'e}sz{\'a}ros}}{{M{\'e}sz{\'a}ros}}{2002}%
]{Meszaros_2002}
{M{\'e}sz{\'a}ros} P.,  2002, \araa, 40, 137

\bibitem[\protect\citeauthoryear{{M{\'e}sz{\'a}ros}}{{M{\'e}sz{\'a}ros}}{2006}%
]{Mezaros1}
{M{\'e}sz{\'a}ros} P.,  2006, Reports on Progress in Physics, 69, 2259

\bibitem[\protect\citeauthoryear{{Nakar} \& {Piran}}{{Nakar} \&
  {Piran}}{2002}]{Nakar_Piran_2002}
{Nakar} E.,  {Piran} T.,  2002, \mnras, 331, 40

\bibitem[\protect\citeauthoryear{{Nishikawa}, {Hardee}, {Richardson}, {Preece},
  {Sol} \& {Fishman}}{{Nishikawa} et~al.}{2003}]{Nishikawa_2003}
{Nishikawa} K.,  {Hardee} P.,  {Richardson} G.,  {Preece} R.,  {Sol} H.,
  {Fishman} G.~J.,  2003, \apj, 595, 555

\bibitem[\protect\citeauthoryear{{Nishikawa}, {Hardee}, {Richardson}, {Preece},
  {Sol} \& {Fishman}}{{Nishikawa} et~al.}{2005}]{Nishikawa_etal_2005}
{Nishikawa} K.,  {Hardee} P.,  {Richardson} G.,  {Preece} R.,  {Sol} H.,
  {Fishman} G.~J.,  2005, \apj, 622, 927

\bibitem[\protect\citeauthoryear{{Nishikawa}, {Niemiec}, {Sol}, {Medvedev},
  {Zhang}, {Nordlund}, {Frederiksen}, {Hardee}, {Mizuno}, {Hartmann} \&
  {Fishman}}{{Nishikawa} et~al.}{2008}]{Nishikawa_etal_2008}
{Nishikawa} K.,  {Niemiec} J.,  {Sol} H.,  {Medvedev} M.,  {Zhang} B.,
  {Nordlund} {\AA}.,  {Frederiksen} J.,  {Hardee} P.,  {Mizuno} Y.,  {Hartmann}
  D.~H.,    {Fishman} G.~J.,  2008, in {F.~A.~Aharonian, W.~Hofmann, \&
  F.~Rieger} ed., American Institute of Physics Conference Series Vol.~1085 of
  American Institute of Physics Conference Series, {New Relativistic
  Particle-In-Cell Simulation Studies of Prompt and Early Afterglows from
  GRBs}.
pp 589--593

\bibitem[\protect\citeauthoryear{{Ohno}, {Fukazawa}, {Takahashi}, {Yamaoka},
  {Sugita} \& {Pal'Shin}}{{Ohno} et~al.}{2008}]{Ohno__Palshin_etal_2008}
{Ohno} M.,  {Fukazawa} Y.,  {Takahashi} T.,  {Yamaoka} K.,  {Sugita} S.,
  {Pal'Shin} V. e.~a.,  2008, \pasj, 60, 361

\bibitem[\protect\citeauthoryear{{Paczynski} \& {Rhoads}}{{Paczynski} \&
  {Rhoads}}{1993}]{Rhoads1993}
{Paczynski} B.,  {Rhoads} J.~E.,  1993, \apjl, 418, L5+

\bibitem[\protect\citeauthoryear{{Pal'shin}, {Aptekar}, {Frederiks},
  {Golenetskii}, {Il'Inskii} \& {Mazets}}{{Pal'shin}
  et~al.}{2008}]{Palshin_etal_2008}
{Pal'shin} V.,  {Aptekar} R.,  {Frederiks} D.,  {Golenetskii} S.,  {Il'Inskii}
  V.,    {Mazets} E. e.~a.,  2008, in {M.~Galassi, D.~Palmer, \& E.~Fenimore}
  ed., American Institute of Physics Conference Series Vol.~1000 of American
  Institute of Physics Conference Series, {Extremely long hard bursts observed
  by Konus-Wind}.
pp 117--120

\bibitem[\protect\citeauthoryear{{Piran}}{{Piran}}{2005}]{Piran_2005}
{Piran} T.,  2005, in AIP Conf. Proc. 784: Magnetic Fields in the Universe:
  From Laboratory and Stars to Primordial Structures. {Magnetic Fields in
  Gamma-Ray Bursts: A Short Overview}.
pp 164--174

\bibitem[\protect\citeauthoryear{{Preece}, {Briggs}, {Mallozzi}, {Pendleton},
  {Paciesas} \& {Band}}{{Preece} et~al.}{2000}]{Precee_2000}
{Preece} R.~D.,  {Briggs} M.~S.,  {Mallozzi} R.~S.,  {Pendleton} G.~N.,
  {Paciesas} W.~S.,    {Band} D.~L.,  2000, \apjs, 126, 19

\bibitem[\protect\citeauthoryear{{Sari}}{{Sari}}{2006}]{Sari_2006}
{Sari} R.,  2006, in {P.~A.~Hughes \& J.~N.~Bregman} ed., Relativistic Jets:
  The Common Physics of AGN, Microquasars, and Gamma-Ray Bursts Vol.~856 of
  American Institute of Physics Conference Series, {Gamma Ray Bursts and Their
  Afterglows}.
pp 33--56

\bibitem[\protect\citeauthoryear{{Silva}}{{Silva}}{2006}]{Silva_2006}
{Silva} L.~O.,  2006, in {Hughes} P.~A.,  {Bregman} J.~N.,  eds, AIP Conf.
  Proc. 856: Relativistic Jets: The Common Physics of AGN, Microquasars, and
  Gamma-Ray Bursts. {Physical Problems (Microphysics) in Relativistic Plasma
  Flows}.
pp 109--128

\bibitem[\protect\citeauthoryear{{Sironi} \& {Spitkovsky}}{{Sironi} \&
  {Spitkovsky}}{2009}]{Sironi_Spitkovsky_2009}
{Sironi} L.,  {Spitkovsky} A.,  2009, ArXiv e-prints

\bibitem[\protect\citeauthoryear{{Toptygin} \& {Fleishman}}{{Toptygin} \&
  {Fleishman}}{1987}]{Topt_Fl_1987}
{Toptygin} I.~N.,  {Fleishman} G.~D.,  1987, \apss, 132, 213

\bibitem[\protect\citeauthoryear{{Workman}, {Morsony}, {Lazzati} \&
  {Medvedev}}{{Workman} et~al.}{2008}]{Workman_etal_2008}
{Workman} J.~C.,  {Morsony} B.~J.,  {Lazzati} D.,    {Medvedev} M.~V.,  2008,
  \mnras, 386, 199

\end{thebibliography}

\label{lastpage}

\end{document}